\documentclass{icrc}
\usepackage{graphicx}
\usepackage{amsmath}
\begin{document}

\title{Anomalous diffusion of the cosmic rays: steady-state
solution}
\author[1]{Lagutin A.A}
\author [2]{Makarov V.V.}
\author[1]{Tyumentsev A.G.}
\affil[1]{Altai State University, Barnaul 656099, Russia}
\affil[2]{Moscow State University,Moscow, Russia}

\correspondence{A.A.Lagutin(lagutin@theory.dcn-asu.ru)}

\firstpage{1}\pubyear{2001}

\maketitle

\begin{abstract}

We consider the propagation of galactic cosmic rays in the fractal
interstellar medium. Steady state solution of the fractional
diffusion equation, describing cosmic ray propagation, is found.
We show that the exponent of the steady state spectrum turns out
to be equal to the exponent of the cosmic ray spectrum above the
``knee'', that is $\eta\approx 3.1$.
\end{abstract}

\section{Introduction}

The steady state diffusion equation is frequently used for
interpretation of the cosmic ray phenomena (see, for
example,\citet{b4,b5,b9,b10,b11,b12}). Without energy losses and
nuclear interactions this equation for concentration of the cosmic
rays with energy $E$ generated by sources distribution with
density function $S(\vec r,E)$ has the form:
\begin{equation}\label{eq1l}
  D(E)\triangle N(\vec r,E)+S(\vec r,E)=0,
\end{equation}
where $D(E)$ is a diffusivity. The equation (\ref{eq1l}) describes
the cosmic ray propagation under the assumption that
nonhomogeneities of medium have small-scale character. If the
medium has a fractal structure, the normal diffusion equation
(\ref{eq1l}) is not a proper one. A generalization of this
equation leads to an anomalous diffusion (see the reviews by
\citet{b1l,b2l,b3l,b7} and paper \citet{b18}). In recent papers
\citep{b1,b8,b18} the anomalous diffusion equations under the
different approximations were formulated and the solutions for
point instantaneous and impulse sources were found.

The main goal of this paper is to find the steady state solution
of the fractional diffusion equation and to evaluate the exponent
of steady state energy spectrum.

\section{Fractional diffusion equation}

The fractional diffusion equation formulated in \citep{b18}
has the form :
\begin{multline} \label{eq1dd}
\frac{\partial N(\vec r,E,t)}{\partial
   t}=-D(E,\alpha,\beta)D_{0+}^{1-\beta}(-\triangle)^{\alpha/2}N(\vec
   r,E,t)\\
 \qquad\qquad +S(\vec r,E,t),
\end{multline}
here $D(E,\alpha,\beta)=D_0(\alpha,\beta)E^{\delta}$ is the
anomalous diffusivity, $\alpha$, $\beta$ are determined by
the fractal structure of the medium and by trapping
mechanism, correspondingly. $D_{0+}^{\mu} $ means
fractional derivative of Rieman-Liouville by
time~\citep{b2}:
\[
 D_{0+}^{\mu}f\equiv\frac{1}{\Gamma(1-\mu)}\frac{d}{dt}
 \int\limits_0^t(t-\tau)^{-\mu}f(\tau)d\tau,\quad \mu < 1,
\]
$(-\triangle)^{\alpha/2}$ is fractional Laplacian (called
``Riss' operator'')~\citep{b2}:
\[
(-\triangle)^{\alpha/2}f(x)=\frac{1}{d_{m,l}(\nu)}
\int\limits_{R^m}\frac{\triangle^l_y f(x)}{|y|^{m+\nu}} dy,
\]
 where  $l > \alpha$, $x \in {\rm R}^m$, $y \in {\rm R}^m$,
 \[
  \Delta_{y}^{l} f(x) = \sum_{k=0}^{l} (-1)^{k} {l\choose
k} f(x-ky)
\]
\[
 d_{m,l}(\nu) =
\int\limits_{{\rm R}^m}(1-e^{iy})^l |y|^{-m-\nu}dy,
\]

If $\alpha=2$ and $\beta=1$, the equation (\ref{eq1dd}) is the
normal diffusion equation. If $\alpha<2$, $\beta=1$, we have from
(\ref{eq1dd}) the superdiffusion equation:
\begin{multline} \label{eq2l}
\frac{\partial N(\vec r,E,t)}{\partial
   t}=-D(E,\alpha)(-\triangle)^{\alpha/2}N(\vec r,E,t) \\
\qquad\qquad +S(\vec r,E,t),
\end{multline}
discussed in \citet{b8}.

\section{Steady state solution of superdiffusion equation}

In steady state case the equation (\ref{eq2l}) takes the form:
\begin{equation}\label{eq2}
  D(-\triangle)^{\alpha/2}N(\vec r,E)=S(\vec r,E).
\end{equation}
 The Green's function $G(\vec r,E;E_0)$ satisfies the equation:
\begin{equation}\label{eq3}
D(-\triangle)^{\alpha/2}G(\vec
r,E;E_0)=\delta(E-E_0)\delta(\vec r).
\end{equation}
The solution of the equation (\ref{eq3}) can be found by means of
Fourier transformation:
\[
  \widetilde{f}(\vec k)=\widehat{F}f(\vec
  r)=\int\limits_{{R}^3}e^{i\vec k\vec r}f(\vec
  r)d\vec r.
\]
Taking into account that the Fourier transform of fractional
Laplacian is \citep{b2}
\[
\widehat{F} (-\triangle )^{\alpha/2}G(\vec r,E;E_0)=|\vec
k|^\alpha\widetilde{G}(\vec k,E;E_0),
\]
it's easy to find:
\[
\widetilde{G}(\vec k,E;E_0)=\frac{\delta(E-E_0)}{D|\vec
k|^\alpha}=\delta(E-E_0) \int\limits_0^{\infty}dye^{-D|\vec
k|^\alpha y}.
\]
Applying the inverse Fourier transformation, we obtain:
\[
G(\vec r,E;E_0)= \frac{\delta(E-E_0)}{(2\pi)^3}
\int\limits_0^{\infty} dy\int\limits_{{R}^3} d\vec k e^{-i\vec
k\vec r-D|\vec k|^{\alpha}y}.
\]
 Since
\[
g^{(\alpha)}_3(r)
=\frac{1}{(2\pi)^3}\int\limits_{{R}^3}d\vec ke^{-i\vec
k\vec r-|\vec k|^\alpha}
\]
 is density of three-dimensional spherically-symmetrical
stable law, for Green's function we have:
\begin{equation}\label{eq12}
G(\vec r,E;E_0)= \frac{\delta(E-E_0)}{(D)^{3/\alpha}}
\int\limits_0^{\infty} dyy^{-3/\alpha} g_3^{(\alpha)}\left(
\frac{|\vec r|}{(Dy)^{1/\alpha}} \right).
\end{equation}

Using Green's function (\ref{eq12}) we can find the solution of
equation (\ref{eq2}) for a interesting for astrophysics source.
Thus, for point source with inverse power spectrum relating to
supernova burst $S(\vec r,E)=S_0E^{-p}\delta(\vec r)$, the
solution of the equation has the form:
\begin{equation}\label{eq13}
  N(\vec r,E)=\frac{S_0E^{-p}}{(D)^{3/\alpha}}\int\limits_0^{\infty}dy
y^{-3/\alpha}g_3^{(\alpha)}\left( \frac{|\vec
r|}{(Dy)^{1/\alpha}} \right)
\end{equation}

Taking into account Mellin transform of spherically-symmetrical
three-dimensional stable laws \citep{b7}:
\begin{equation}\label{eq23}
  g_m^{(\alpha)}(s)=\int\limits_0^\infty
  g_m^{(\alpha)}(r)r^{s-1}dr=\frac{2^s\Gamma\left(\frac{s}{2}\right)
  \Gamma\left(\frac{m-s}{\alpha}\right)}{\alpha(4\pi)^{m/2}
   \Gamma\left(\frac{m-s}{2}\right)},
\end{equation}
we have
\begin{equation}\label{eq27}
  N(\vec r,E)=
  \frac
   {2^{-\alpha}S_0}
     {\pi ^{3/2}D_0r^{3-\alpha}}
   \frac
     {\Gamma\left(\frac{3-\alpha}{2}\right)}
      {\Gamma\left(\frac{\alpha}{2}\right)}
  E^{-p-\delta}.
\end{equation}

\begin{figure}[t]
\vspace*{2.0mm} 
\includegraphics[width=8.3cm]{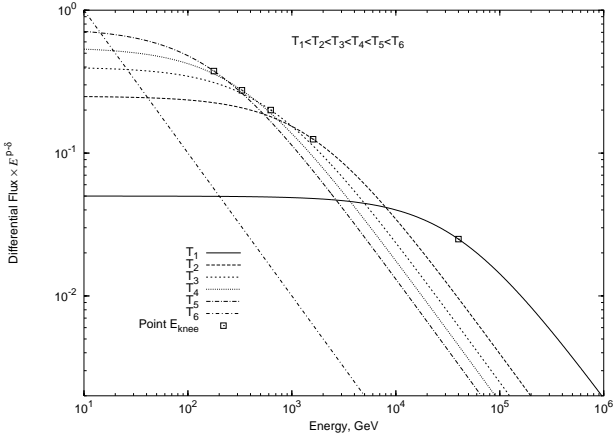} 
\caption{Modification of the energy spectrum versus $T$
($T_6\rightarrow\infty$). Square-position of the
``knee''}\label{f1}
\end{figure}

To evaluate numerically the value of spectral exponent
$\eta=p+\delta$, let's consider in details a passage from
non-stationary solution to steady state one in a special case
$\alpha=1$. This choice is due to the fact that $g_3^{(1)}$ has
the analytical representation. It's three-dimensional Cauchy's
density:
\[
g^{(1)}_3=\frac{1}{\pi^2(1+r^2)^2}.
\]
Based on the results obtained in \citet{b8} for points
impulse source
\[
S({\vec r},t,E)=S_0 E^{-p} \delta({\vec r})
\Theta(T-t)\Theta(t),
\]
in our case we have
\begin{equation}\label{eq3l}
N(\vec r,T,E)=\frac{S_0E^{-p-\delta}}{2\pi D_0r^2T}
\left[1-\frac{1}{\frac{D_0^2T^2}{r^2}E^{2\delta}+1}\right].
\end{equation}
Using the representation $N(\vec r,E)=N_0E^{-\eta}$ from
(\ref{eq3l}), one can easily find the spectral exponent:
\begin{equation}\label{eq21}
  \eta(T)=p+\delta-\frac{2\delta}{\left(\frac{TD_0E^\delta}{r}+1\right)}
\end{equation}
It should be noted that the solution (\ref{eq3l}) has the ``knee''
at $E_{0}(T)=\left(\frac{r}{TD_0}\right)^{1/\delta}$. At the
``knee'' energy $E_0$ the spectral exponent for observed particles
$\eta$ is equal to the spectral exponent for particles generated
by the sources:
\[
\eta=p.
\]
One can also see from (\ref{eq21}) that at $E\ll E_0$ and $E\gg
E_0$, we have correspondingly:
\[
\eta_{E\ll E_0}\approx p-\delta,
\]
\begin{equation}\label{eq5l}
\eta_{E\gg E_0}\approx p+\delta,
\end{equation}
that is the spectrum steepening on $2\delta$. The steady state
solution is connected with $N(\vec r,T,E)$ by means of passage to
the limit:
\begin{equation}\label{eq6l}
N(\vec r,E)=\lim\limits_{T\rightarrow\infty}N(\vec r,T,E).
\end{equation}
So, we have
\begin{equation}\label{eq7l}
\eta=\lim\limits_{T\rightarrow\infty}\eta(T)=p+\delta.
\end{equation}
\begin{equation}\label{eq8l}
E_0=\lim\limits_{T\rightarrow\infty}E_0(T)\rightarrow 0.
\end{equation}
Thus, the analysis presented above shows that the exponent of the
steady state spectrum turns out to be equal to the spectral
exponent above the ``knee'', that is $p+\delta\approx 3.1$. This
conclusion is illustrated by (fig.\ref{f1}).

\section{Steady state solution of fractional \\ diffusion equation($\alpha<2, \beta<1$)}
It has been shown in \citep{b18} that the solution of
equation (\ref{eq1dd}) for the point impulse source has the
form:
\begin{multline}
N(\vec r,t,E)=\frac{S_0E^{-p}}{D(E,\alpha,\beta)^{3/\alpha}}
\int\limits^t_{max[0,t-T]}\tau^{-3\beta/\alpha} \\
\times\Psi_3^{(\alpha,\beta)}\left(|\vec
r|(D(E,\alpha,\beta)\tau^{\beta})^{-1/\alpha}\right),
\end{multline}
where the scaling function $\Psi_3^{(\alpha,\beta)}(r)$,
\begin{equation}\label{eq4dd}
  \Psi_3^{(\alpha,\beta)}\left(r\right)=\int\limits_0^{\infty}
  q_3^{(\alpha)}(r\tau^{\beta})
  q_1^{(1,\beta)}(\tau)\tau^{3\beta/\alpha}d\tau,
\end{equation}
 is determined by three-dimensional spherically-symmetrical
stable distribution $q_3^{(\alpha)}(r)$ and
 one-sided stable
distribution $q_1^{(1,\beta)}(t)$ with characteristic
exponent $\beta<1$~\citep{b7}:
\begin{equation}\label{eq7dd}
q_1^{(\beta,1)}(t)=(2\pi
i)^{-1}\int\limits_{\gamma-i\infty}^{\gamma+i\infty} e^{(\lambda
t-\lambda^{\beta})}d\lambda.
\end{equation}

As the integral diverges if $T\rightarrow\infty$, the
spectral exponent has been evaluated for $T\sim 10^{10}y$.
We found $\eta\approx p+\frac{\delta}{\beta}$.

\balance
\section{Conclusion}

We have considered the propagation of galactic cosmic rays in the
fractal interstellar medium. Steady state solution of the
fractional diffusion equations describing cosmic ray propagation
have been found. We have shown that the exponent of the steady
state spectrum turns out to be equal to the exponent of cosmic ray
spectrum above the ``knee'', that is $\eta\approx 3.1$.


\begin{thebibliography}{99}
\bibitem[Berezinsky et al.(1990)]{b5}Berezinsky V.S., Bulanov S.V., Ginzburg V.L. et al. Astrophysics of cosmic
rays. North Holland, Amsterdam, 1990.
\bibitem[Blumen et al.(1991)]{b9}Blumen I.B.G.M., Dogel V.A., Dorman V.L., Ptuskin V.S.,
Izv.AN SSSR, Ser.phys., 55, 2052-2055, 1991.
\bibitem[Bouchaud and Georges (1990)]{b1l}Bouchaud J.-P.,
Georges A. Phys.Rep., 195, 127-293, 1990.
\bibitem[Ginzburg and  Syrovatskii(1964)]{b4}Ginzburg V.L., Syrovatskii S.I. Origin of cosmic rays. Pergamon Press, 1964.
\bibitem[Isichenko(1992)]{b2l}Isichenko M.B. Rev.Mod.Phys.
64, 961-1043, 1992.
\bibitem[Kalmykov et al(1999)]{b12}Kalmykov N.N., Pavlov A.I.,Proc.25 ICRC,
 4, 293-296, 1999.
\bibitem[Lagutin et al.(2000)]{b1}Lagutin A.A., Nikulin Yu.A. Uchaikin V.V. Preprint ASU--2000/4, Barnaul,
2000.
\bibitem[Lagutin et al.(2001)]{b8}Lagutin A.A., Nikulin Yu.A., Uchaikin V.V.,Nuclear Physics B (Proc.Suppl.)
97, 267-270, 2001.
\bibitem[Lagutin and Uchaikin(2001)]{b18}
Lagutin A.A. and Uchaikin V.V. Fractional diffusion of cosmic
rays. Proceedings of the 27th ICRC, 2001.
\bibitem[Ptuskin et al.(1993)]{b11} Ptuskin V.S., Rogovaya S.I., V.N.Zirakashvili at
al., Astron.Astrophys., 268, 726-735, 1993.
\bibitem[Samko et al.(1987)]{b2}Samko S.G., Kilbas A.A., Marichev O.I. Fractional integrals and derivations
and some applications. Minsk: Nauka, 1987 (in Russian).
\bibitem[Uchaikin and Zolotarev(1999)]{b7}Uchaikin V.V., Zolotarev V.M. Chance and Stability. VSP, Netherlands,
  Utrecht, 1999.
\bibitem[West end Deering(1994)]{b3l}West B.J.,Deering W.
Phys. Rep., 246, 1-100, 1994.
\bibitem[Zirakashvili et al.(1991)]{b10}Zirakashvili V.N., Klenach E.G., Ptuskin V.S.,
Izv. AN SSSR, Ser. phys., 55, 2049-2051, 1991.
\end{thebibliography}
\end{document}